# Deflection of coronal rays by remote CMEs: shock wave or magnetic pressure?

Boris Filippov[1] and A.K. Srivastava[2]

[1]*Pushkov Institute of Terrestrial Magnetism, Ionosphere and Radio Wave Propagation, Russian Academy of Sciences, Troitsk Moscow Region 142190 Russia*
*(e-mail:* bfilip@izmiran.troitsk.ru*)*

[2]*Aryabhatta Research Institute of Observational Sciences (ARIES), Manora Peak, Nainital-263 129, Uttarakhand, India.*

*(e-mail:* aks@aries.res.in*)*

**Abstract.** We analyze five events of the interaction of coronal mass ejections (CMEs) with the remote coronal rays located up to 90º away from the CME as observed by the SOHO/LASCO C2 coronagraph. Using sequences of SOHO/LASCO C2 images, we estimate the kink propagation in the coronal rays during their interaction with the corresponding CMEs ranging from 180 to 920 km s$^{-1}$ within the interval of radial distances form 3 $R_\odot$ to 6 $R_\odot$. We conclude that all studied events do not correspond to the expected pattern of shock wave propagation in the corona. Coronal ray deflection can be interpreted as the influence of the magnetic field of a moving flux rope related to a CME. The motion of a large-scale flux rope away from the Sun creates changes in the structure of surrounding field lines, which are similar to the kink propagation along coronal rays. The retardation of the potential should be taken into account since the flux rope moves at high speed comparable with the Alfvén speed.

**Keywords:** Coronal mass ejections (CMEs); Coronal rays; Streamers; Magnetic fields; Shock waves; Kink waves

## 1. Introduction

Coronal mass ejections (CMEs) are the most violent and fast propagating disturbances of the million degree hot outer solar corona. They drastically change the overall view of the corona in the field of view of a spaceborne coronagraph, and influence nearby coronal structures as well as remote coronal rays. Halo CMEs reveal some changes in the coronal brightness all over the solar limb. However, this is the sky plane projection of a structure moving close to the line of sight, whose angular size is greater than the occulting disk of a coronagraph. But some coronal structures show indubitable physical action of a remote CME. Thin coronal rays located several tens of degrees away from a propagating CME are deflected and become temporarily curved. The hump of the ray moves along it outwards as a kink perturbation. Many authors interpret this pattern as a manifestation of a shock produced by fast, super-Alfvenic CME propagation in the corona (e.g., Hundhausen, 1987; Sime and Hundhausen, 1987; Sheeley, Hakala, and Wang, 2000; van der Holst, van Driel-Gesztelyi, and Poedts, 2002; Tripathi and Raouafi, 2007; Vourlidas and Ontiveros, 2009, and references cited there).

However, noting that only a small fraction of observed CMEs moving faster than the Alfvén speed (600 km s$^{-1}$ at 3 $R_\odot$), Hundhausen (1987) concluded that relatively few CMEs would be able to produce deflections of remote streamers by driving shocks through the corona. Sime and Hundhausen (1987) found only one event for which a shock may be the most likely candidate for the deflection of the coronal rays during the propagation of the CMEs. Hundhausen (1987) reasonably explained that most of the streamer deflections were caused by compressive magnetoacoustic waves moving out from the sides of CMEs approximately transverse to the nearly radial magnetic field. Sheeley, Hakala, and Wang (2000) found that kink disturbances decelerate as they move radially outward along the rays. They interpreted this as an indication of the slowing down of the shock wave front as it is pushed obliquely across the radial magnetic field lines indicated by the deflected streamers and coronal rays. On the other hand, Ajabshirizadeh and Filippov (2004) have suggested that the magnetic field of a moving CME deflects streamer plasma from its straight radial flow. In the frame of the concept that most

CMEs are created by flux rope eruptions, a CME carries rather strong electric current which is a source of additional coronal magnetic field. When the CME reaches the outer corona, the role of this source becomes more important for the surrounding coronal structures. The magnetic pressure of the CME current is able to influence a vast coronal volume and change the structure of ambient magnetic field and plasma flows.

In this paper, we analyze the deflection of coronal rays by remote CMEs and discuss the possibility of interpreting these changes as the influence of the moving CME electric current on the coronal structure. In section 2, we present the observational data. We describe the magnetic field model in section 3. In section 4, we present results and discussion. We discuss the conclusions in the last section.

## 2. Observational Data

We analyze five events involving interaction of CMEs with the remote coronal rays as observed by the SOHO/LASCO C2 coronagraph. All five of these events show prominent coronal ray deflections caused by the observed CMEs. In each event described in our paper, we use corresponding SOHO/LASCO C2 temporal image data to measure the speed of kink propagation along the observed coronal rays that are deflected by propagating CMEs. Some of these events were considered by other authors, and we chose them in order to compare our results with the results of those previous works.

### *Event of 22 August 1996*

According to the SOHO LASCO CME Catalog (http://cdaw.gsfc.nasa.gov/CME_list/), the coronal mass ejection (CME) appeared in the field of view of the C2 coronagraph at 08:39 UT on 22$^{nd}$ August 1996. It moved in the south-east (SE) direction (the position angle $P$ was about 130°) at a constant speed of about 1000 km s$^{-1}$ in the field of view of the C2 coronagraph. In the present paper, henceforth, we use the term 'speed' for the apparent velocity of a structure in the sky plane. The width of the CME was more than 100° and three prominent streamer rays were also located above the north-east (NE) limb within the interval of position angles 40° - 80°.

In order to better see the tangential displacement of these coronal rays, we construct a map from the strips of C2 images oriented nearly perpendicular to the rays. We have arranged these time slices in the ascending time sequence of the observation. Figure 2 shows two slice-time diagrams respectively for the heliocentric distances 3.5 $R_\odot$ and, 4.9 $R_\odot$. The observation time is in ascending order from left to right in both the panels of Figure 2. Displacement curves are presented in Figure 3 for the three heliocentric distances 3.5 $R_\odot$, 4.9 $R_\odot$, and 6.3 $R_\odot$ respectively. One such curve is also over-plotted in the left panel of Figure 2 for the heliocentric distance 3.5 $R_\odot$. To measure the speed of the kink propagation, we follow in time the behavior of some characteristic points of the hump. This may be the initiation point of the deflection which is difficult to trace, or the maximum deviation point. We fix the moments of the maximum deflection at the three distances as well as the moments of a noticeable displacement of streamer axes at one third height of these curves. The speed of the front of curves $v_{f1}$ within the spatial interval 3.5 $R_\odot$ – 4.9 $R_\odot$ and the speed $v_{f2}$ within the spatial interval 4.9 $R_\odot$ – 6.3 $R_\odot$ are shown in the Table 1. The average speed $v_f$ is about 630 km s$^{-1}$. However, as the front shifts only to the next slice in the next map, the error can be of the order of 50%. The right three columns of Table 1 represent the speed of the curve peak, which is lower and only reaches an average speed of 380 km s$^{-1}$.

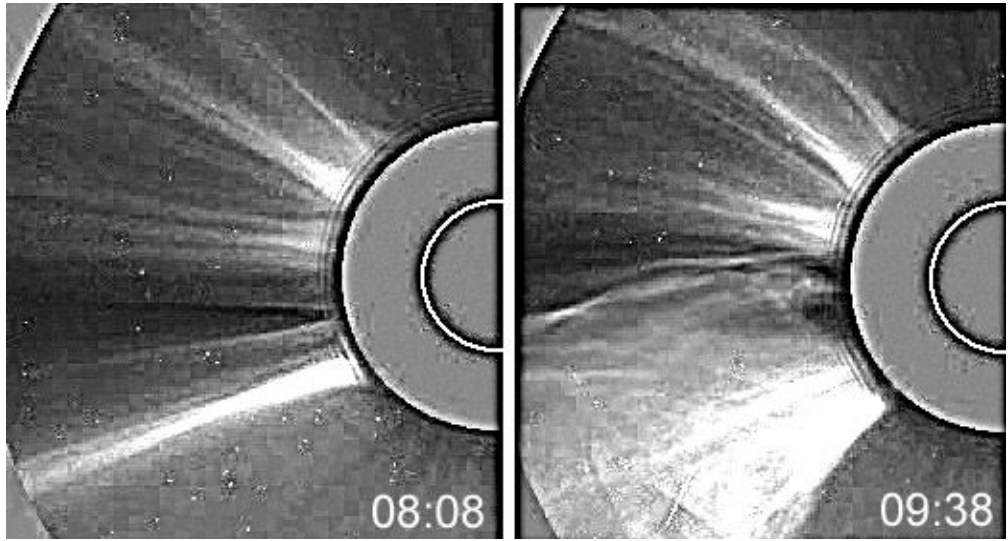

*Figure 1.* SOHO/LASCO C2 images showing deflection and bending of coronal rays during CME propagation on 22 August 1996 (Courtesy:- SOHO/LASCO).

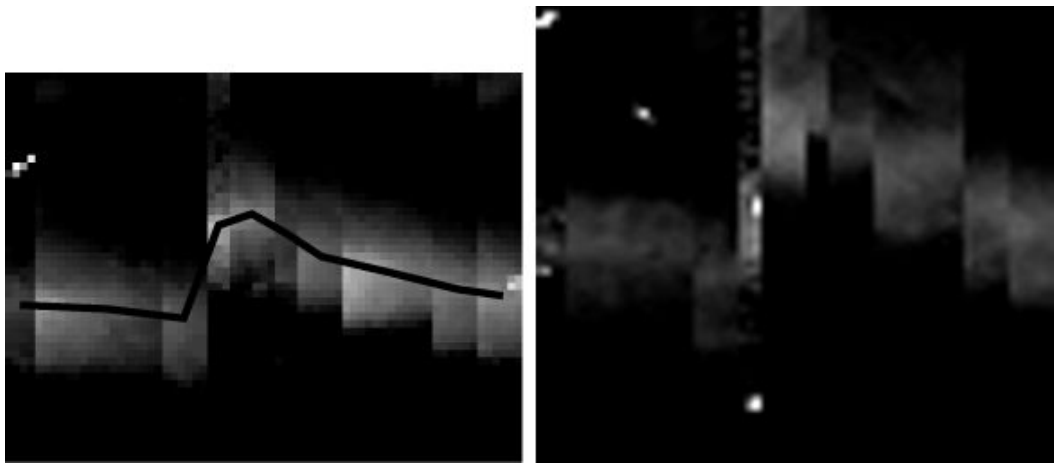

*Figure 2.* Slice-time diagrams for two distances from the Sun for the time interval from 04:36 UT to 15:08 UT on 22 August 1996. Left and right panels correspond to the heliocentric distances of 3.5 $R_\odot$ and 4.9 $R_\odot$ respectively. Observation time is in ascending order from left to the right in each image panel. The black line connects the centroids of ray brightness in each time slice at heliocentric distance 3.5 $R_\odot$.

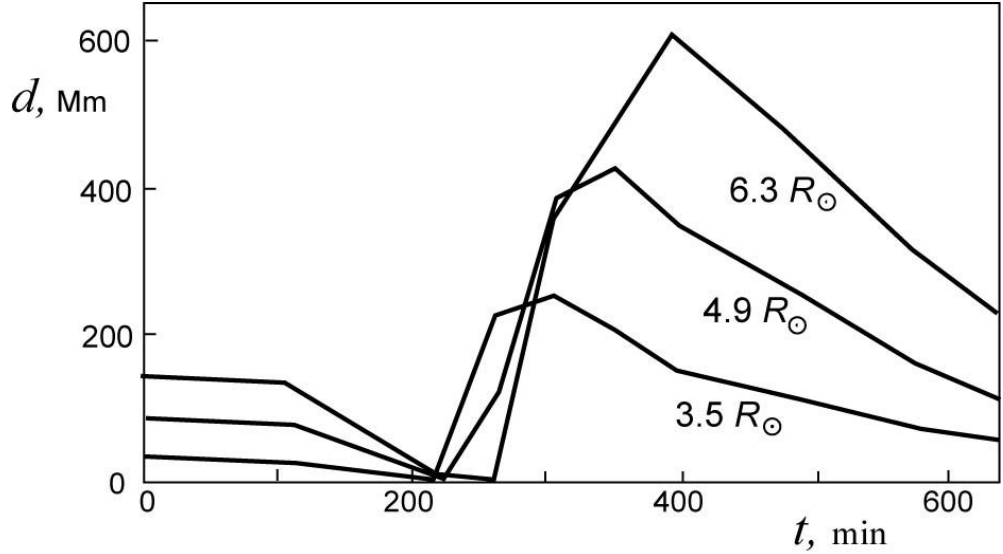

*Figure 3.* Time profiles of the deflection of central coronal ray as shown in Figure 1 and Figure 2. The time origin corresponds to 04:36 UT. The three deflection curves correspond respectively to the heliocentric distances 3.5 $R_\odot$, 4.9 $R_\odot$, and 6.3 $R_\odot$.

TABLE 1
Velocity of kink propagation along streamers

| Event | $v_{f1}$ km s$^{-1}$ | $v_{f2}$ km s$^{-1}$ | $v_f$ km s$^{-1}$ | $v_{M1}$ km s$^{-1}$ | $v_{M2}$ km s$^{-1}$ | $v_M$ km s$^{-1}$ | $v_{CMEf}$ km s$^{-1}$ | $v_{CMEc}$ km s$^{-1}$ |
|---|---|---|---|---|---|---|---|---|
| 22 August 1996 | 520 | 740 | **630** | 370 | 400 | **380** | 1000 | |
| 5 March 2000 | 410 | 270 | **340** | 510 | 260 | **380** | 850 | 200 |
| 13 January 2002 | | | **250** | | | **180** | 250 | 180 |
| 2 March 2002 Southern | 540 | 660 | **600** | 950 | 900 | **920** | 1100 | 800 |
| 2 March 2002 Northern | 580 | 580 | **580** | 640 | 300 | **470** | 1100 | 800 |

### *Event of 5 March 2000*

The CME appeared in the field of view of the C2 coronagraph at 16:54 UT on 5$^{th}$ March 2000. It moved directly to the North at a constant speed of about 850 km s$^{-1}$. The width of the CME was about 60°, and it had a prominent bright core below a compact oval dark cavity. The speed of the core was only 200 km s$^{-1}$, while the speed of the cavity upper boundary was 400 km s$^{-1}$. A double streamer ray was located near 30° to the East from the CME central pass line. Figure 4 shows the displacements of the ray at the heliocentric distances 3.2 $R_\odot$, 4.6 $R_\odot$, and 6.0 $R_\odot$. The speed of the kink propagation is about 380 km s$^{-1}$.

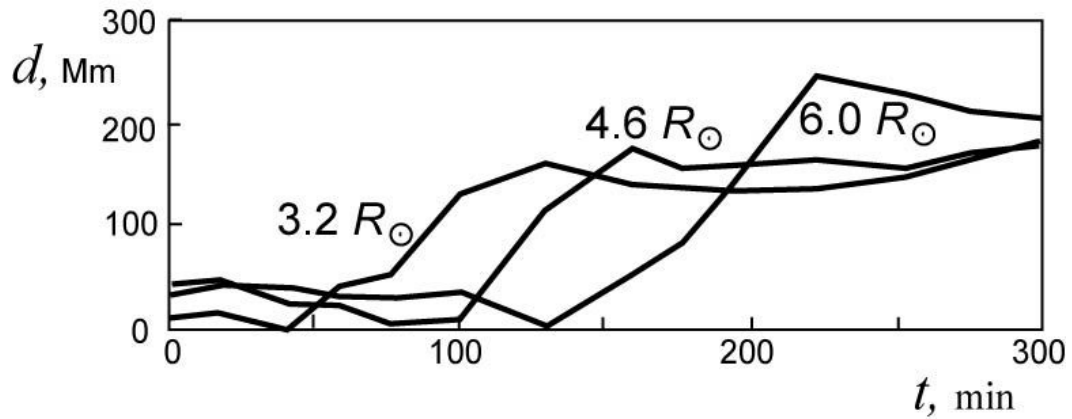

*Figure 4.* Time profiles of the ray deflection on 5 March 2000. Time origin corresponds to 16:54 UT. The three deflection curves correspond respectively to the heliocentric distances 3.2 $R_\odot$, 4.6 $R_\odot$, and 6.0 $R_\odot$.

*Event of 13 January 2002*

A slow CME appeared in the field of view of the C2 coronagraph after 06:25 UT on 13$^{th}$ January 2002. The position angle $P$ of the CME centroid was near 50°. The CME showed noticeable acceleration. The velocity of the leading edge on the half-way in the field of view of the C2 was 250 km s$^{-1}$, while the speed of the bright core was only 180 km s$^{-1}$. The streamer that was deflected by the CME, was located at $P = 30°$. Figure 5 shows the displacements of the ray at the heliocentric distances 3.7 $R_\odot$, and 5.0 $R_\odot$. The hump top moved along the ray at a speed of 180 km s$^{-1}$.

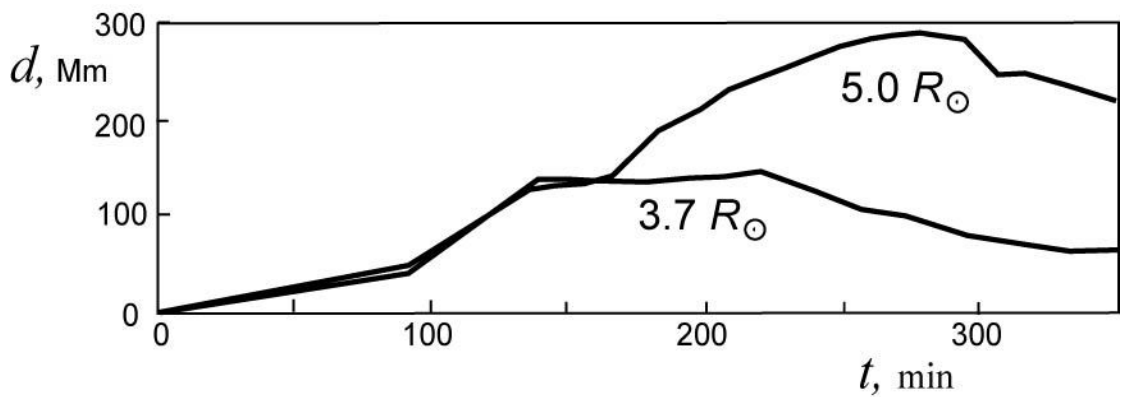

*Figure 5.* Time profiles of the ray deflection on 13 January 2002. Time origin corresponds to 08:40 UT. The two deflection curves correspond respectively to the heliocentric distances 3.7 $R_\odot$ and 5.0 $R_\odot$.

*Event of 2 March 2002*

The wide CME (~ 150°) appeared in the field of view of the C2 coronagraph at 15:06 UT on 2$^{nd}$ March 2002, and moved in the SE direction ($P \sim 110°$) at a constant speed of about 1100 km s$^{-1}$. Tangled threads of the bright core of this CME had also the high speed of ~ 800 km s$^{-1}$. One streamer ray was located at $P = 190°$ or nearly 80° to the South from the CME central pass line. Another streamer was located nearly diametrically opposite the first one, very close to the North pole. Both rays showed bending during the CME passing. Figure 6 shows the displacements of the Southern ray at the heliocentric distances 3.2 $R_\odot$, 4.5 $R_\odot$, and 5.8 $R_\odot$. The average speed of the kink propagation is 920 km s$^{-1}$.

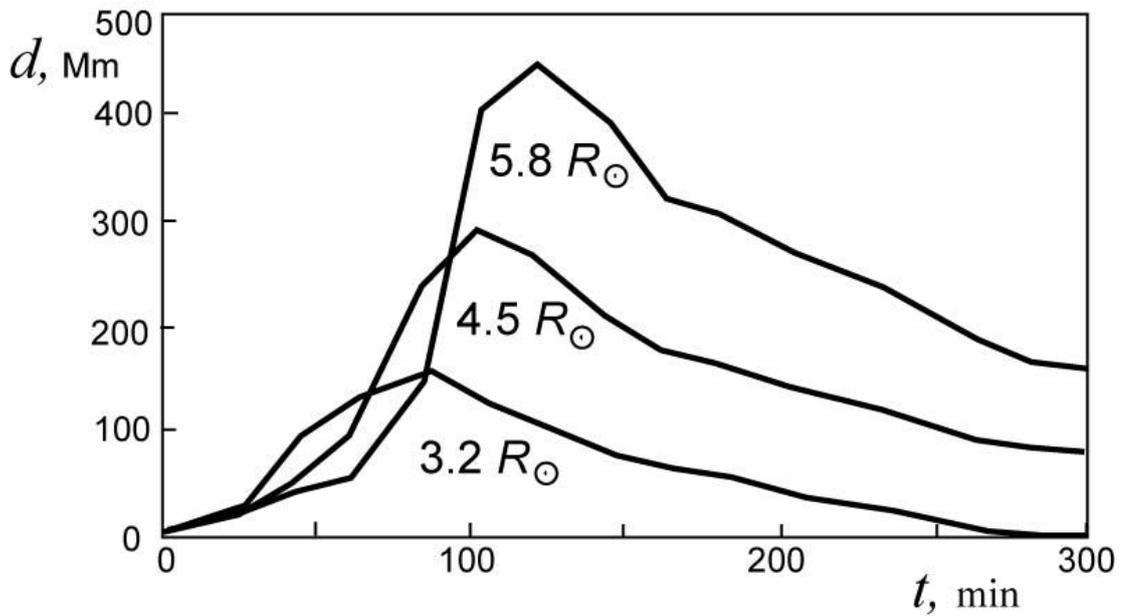

*Figure 6.* Time profiles of the Southern ray deflection on 2 March 2002. Time origin corresponds to 15:06 UT. The three deflection curves correspond respectively to the heliocentric distances 3.2 $R_\odot$, 4.5 $R_\odot$, and 5.8 $R_\odot$.

Figure 7 shows the displacements of the Northern ray at the heliocentric distances 3.6 $R_\odot$, 4.6 $R_\odot$, and 5.6 $R_\odot$. The average speed of the kink propagation is 470 km s$^{-1}$.

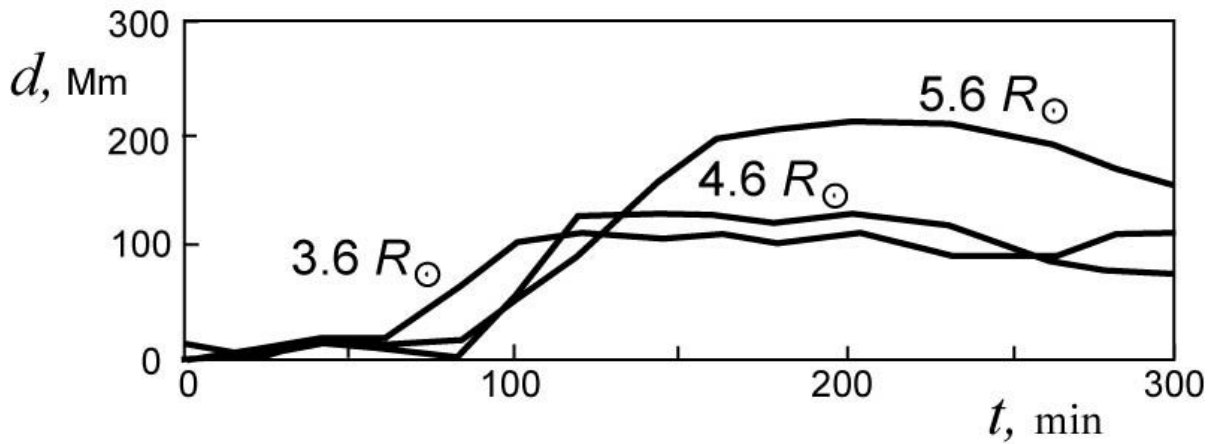

*Figure 7.* Time profiles of the Northern ray deflection on 2 March 2002. Time origin corresponds to 15:06 UT. The three deflection curves correspond respectively to the heliocentric distances 3.6 $R_\odot$, 4.6 $R_\odot$, and 5.6 $R_\odot$.

### 3. Magnetic Field Model

In this section, we consider the changes in the structure of background magnetic field caused by an electric current. In the simplest case, this is a straight linear current in a uniform vertical field approximately corresponding to the solar radial field. Closed circular field lines of the current represent the CME internal magnetic structure. Obviously, field lines become denser and convex on the one side of the current where the current's magnetic field has the same direction as the background field, and they become rarefied and concave on the other side, where the current's magnetic field is opposite to the background field (Figure 8, left panel).

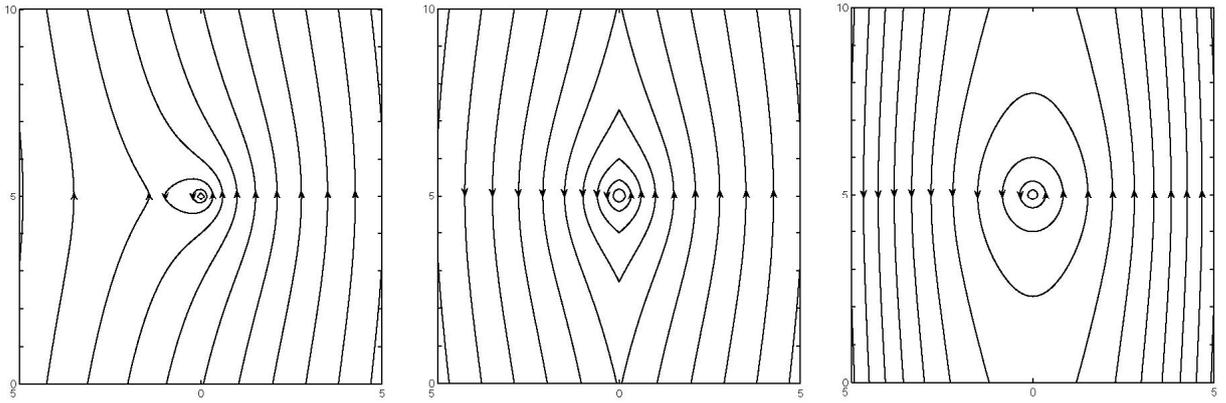

*Figure 8.* Field lines of a linear electric current in the uniform magnetic field (left), within a current sheet (middle), and in the magnetic field with the uniform current density (right).

However, coronal images never show structures concaved away from a nearby CME at the side of it. Concave-outward structures are observed sometimes before or in the frontal part of a CME and are assumed to be associated with slow-mode shocks (Steinolfson and Hundhausen, 1990; Liu *et al.*, 2009). In principle, our model cannot reproduce slow shock formation because plasma pressure is not taken into account. Moreover, if a ring electric current moves with super-Alfvénic velocity as it is shown in Figure 12, the formation of a convex frontal structure due to the slow shock is impossible because the CME speed should be above the sound speed but below the Alfvén speed (Steinolfson and Hundhausen, 1990).

It is believed that CMEs originate from the close vicinity of polarity inversion lines and move nearly along neutral surfaces (Forbes, 2000; Filippov, Gopalswamy, and Lozhechkin., 2001). Therefore, the surrounding magnetic field should be of the opposite directions on different sides of a CME. This condition is satisfied if the current is located within a current sheet, although the field lines have sharp angles in this case (Figure 8, middle). Smoother field lines can be obtained for electric current in the magnetic field with the uniform current density (Figure 8, right). The latter field pattern resembles the structure of a CME and curved coronal rays on both sides of it.

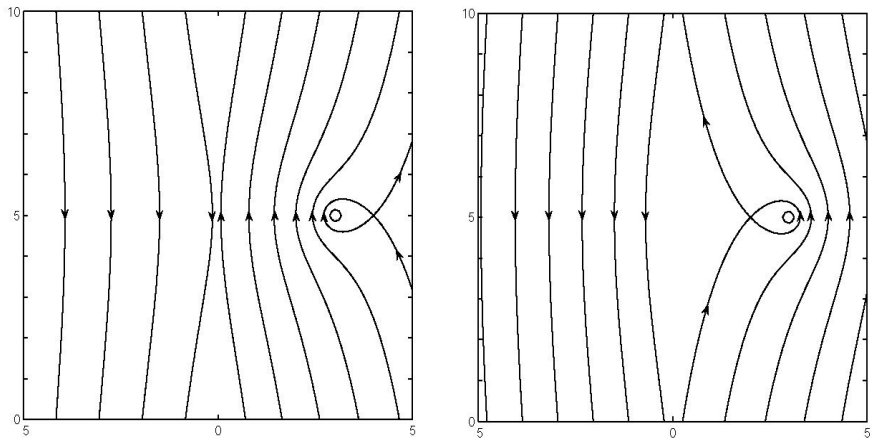

*Figure 9.* Field lines of linear electric currents of opposite directions in the magnetic field with a current sheet.

It should be noted that we assume deflecting coronal rays as structures stretched along the nearly uniform magnetic field. A streamer containing a current sheet inside (Pneuman and Kopp, 1970) will not be deflected by the nearby electric current because field lines on either side of the current sheet are anti-parallel and deviate in opposite directions (Figure 9).

As the current moves along the field lines, the hump or kink of the nearby field lines moves along with it at the same speed. It resembles a kink perturbation in the magnetized plasma. However, it is not a wave but a response of background magnetic field to an additionally moving source of the magnetic field. To be closer to an observed pattern in the field of view of space-borne coronagraphs (e.g., C2, C3 of SOHO/LASCO), we consider a simple 3-D axially symmetric model with the radial magnetic field. The vector potential of the radial field in cylindrical coordinates $\rho, \varphi, z$ can be described by the function

$$A^r_\varphi = M \frac{z^2}{\rho^3}, \qquad (1)$$

where $M$ is the coefficient determining the field strength. A flux rope related to a CME is represented by a ring current $I$ lying in the equatorial plane above the photospheric polarity inversion line. A similar model was analyzed by Lin *et al.* (1998) and Filippov, Gopalswamy, and Lozhechkin (2001). The vector potential of the magnetic field generated by a current ring of the radius $\rho_i$ is given by (Landau, Lifshitz, and Pitaevsky, 1984)

$$A^i_\varphi = \frac{4I}{c\,k} \sqrt{\frac{\rho_i}{\rho}} \left[ (1 - \frac{1}{2}k^2)\,K(k) - E(k) \right], \qquad (2)$$

where

$$k^2 = \frac{4\rho_i \rho}{(\rho_i + \rho)^2 + z^2}, \qquad (3)$$

and $K(k)$, $E(k)$ are the complete elliptic integrals of the first and second kind respectively.

The photospheric plasma is dense and is a good conductor so that it prevents the penetration of the magnetic fields due to the coronal current into the Sun's interior. The surface of the photosphere serves as a spherical mirror for the ring magnetic field. This is equivalent to existence of an additional ring current within the solar interior of the radius $\rho_m$ and strength $I_m$ (Filippov, Gopalswamy, and Lozhechkin, 2001)

$$\rho_m = \frac{R^2}{\rho_i}, \qquad I_m = -I \frac{\rho_i}{R}. \qquad (4)$$

The influence of the mirror current is important for the initial equilibrium of the flux rope but is negligible for the outer corona where deflected streamers are observed. We do not consider here equilibrium condition and equations of motion at all and assume simply the radial motion of the current ring with a constant speed $v_i$ in the outer corona. For initial equilibrium, the $B_z$ component (axis $z$ is perpendicular to the equatorial plane) of the coronal magnetic field in the equatorial plane is necessary, which is absent in the field described by Equation (1).

We take into account the changes of the current due to inductance. It has the form of conservation of the magnetic flux through the ring:

$$LI = \text{const}, \qquad (5)$$

where the self-inductance of the ring with the cross-section radius *a* (Landau, Lifshitz, and Pitaevsky, 1984)

$$L = 4\pi \rho_i \left( \ln \frac{8\rho_i}{a} - \frac{7}{4} \right). \quad (6)$$

For the currents moving at high speed, it is necessary to take into account the retardation of a potential (Landau and Lifshitz, 1998)

$$\mathbf{A} = \frac{1}{c} \int \frac{\mathbf{j}(t - \frac{r}{v_a})}{r} dV, \quad (7)$$

where $v_a$ is the propagation speed of magnetic field changes. In the solar corona it is of the order of Alfven speed

$$v_a = \frac{B}{\sqrt{4\pi \mu n}}. \quad (8)$$

As we have only one moving current *I* and can neglect changes of the mirror current in the present case, we should substitute into Equation (2) the value of $\rho_i$ at the retarded time for any point ($\rho$, $z$) expressed as

$$\rho_i(\rho,z,t) = \frac{k^2(\rho_0 + v_c t) - \rho - \sqrt{k^2(\rho - \rho_0 - v_c t)^2 + (k^2 - 1)z^2}}{k^2 - 1}, \quad (9)$$

where $k = v_a/v_c$ and $\rho_0$ is the initial radius of the ring. Equation (9) is applicable for any $\rho$ and $z$, if $k > 1$, or $v_a > v_c$. In the case $k < 1$, it is applicable only for $\rho < \rho_0 + v_c t$ and $z < k(\rho - \rho_0 + v_c t)/(1-k^2)^{1/2}$. In all other points of space, the magnetic field does not change, and $\rho_i = \rho_0$.

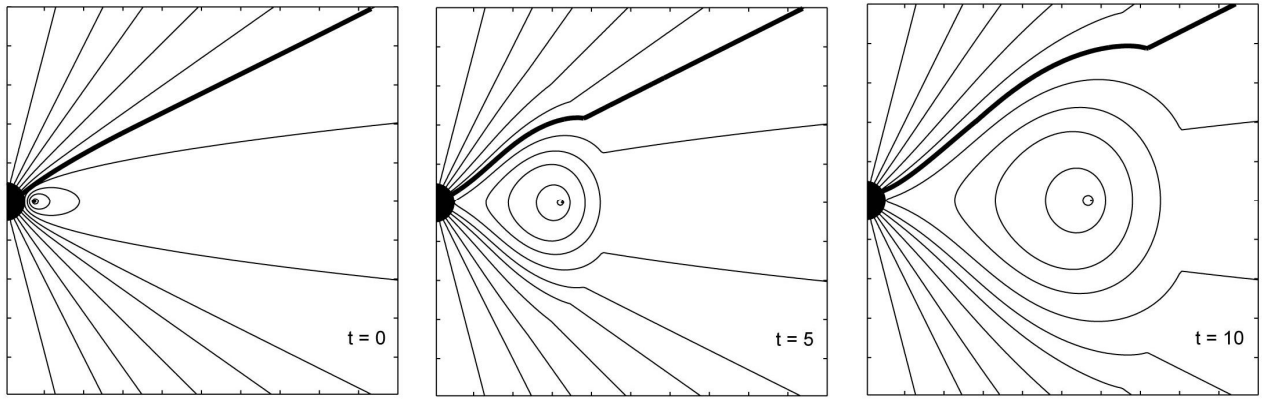

*Figure 10.* Field lines of a ring electric current moving with the velocity $v_c = 0.75\, v_a$ in the radial magnetic field (1).

Figure 10 shows field lines of a ring electric current moving with the velocity $v_c = 0.75\, v_a$ in the radial magnetic field (cf., Eq. 1). The equation of the field lines is

$$\rho(A^r_\varphi + A^i_\varphi) = \text{const}. \quad (10)$$

When all spatial values were expressed in units of the solar radius, the relationship between $I$ and $M$ was chosen as $I/cM = 0.7$ in order to obtain rather wide CME structure. A volume filled with closed field lines (i.e, the CME's body) spreads as it moves away from the Sun into regions of weaker background field, although the strength of the current also decreases according to Equations (5) and (6). Humps on the field lines on both sides of the CME run along the field lines at the speed about $v_c$. Their fronts move at the speed $v_a$ as well as the frontier of closed field lines representing the frontal part of the CME. It looks like the propagation of kink perturbations and the behavior of these field lines is very similar to the behavior of remote coronal rays during a CME event. Figure 11 shows the displacements of the field line marked as a thick line in Figure 10 at the heliocentric distances 3.3 $R_\odot$, 4.8 $R_\odot$, and 6.3 $R_\odot$. The pattern is similar to the plots obtained for the coronal rays in the previous section. Measuring $v_f$ and $v_M$ from the curve shifts in Figure 11, we obtain $v_f = 0.9\ v_a$ and $v_M = 0.8\ v_c$. Thus, time profiles of the deflection give us understated estimates of the velocity of kink propagation seemingly due to the increase of the scale of the hump moving away from the Sun. Nevertheless, $v_f$ and $v_M$ can be used as rather good estimates of $v_a$ and $v_c$ obtained from coronal ray deflection in view of the low cadence of coronal images.

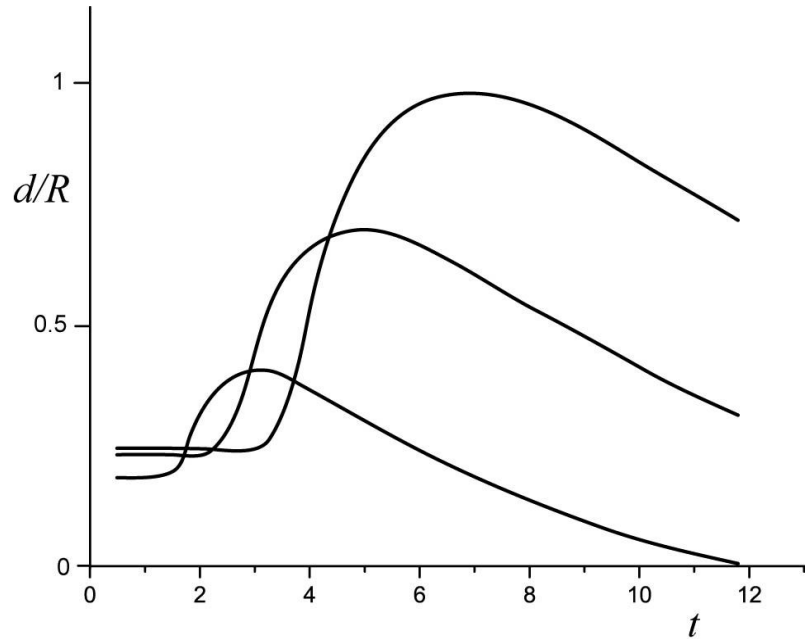

*Figure 11.* Time profiles of the deflection of the field line marked as a thick line in Figure 9 at the heliocentric distances 3.3 $R_\odot$, 4.8 $R_\odot$, and 6.3 $R_\odot$.

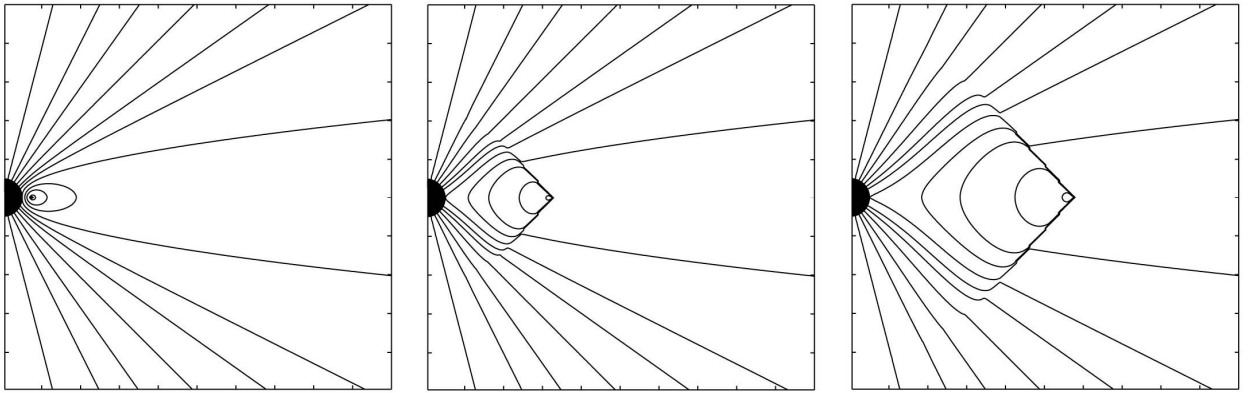

*Figure 12.* Field lines of a ring electric current moving with the velocity $v_c = 1.5\ v_a$ in the radial magnetic field (1).

In the case $v_a < v_c$, we have the field line patterns shown in Figure 12. The super-Alfvénic velocity of the current ring leads to the shock formation. The kinks also propagate along the field lines at about the speed $v_a$. Their fronts form the Mach cone moving before the current ring. The vertex of the cone is sharp because we consider rather thin current ring ($\rho_i /a \gg 1$). If a moving current system fills significant volume, the vertex of the cone will be smoother.

### 4. Results and Discussion

In principle, it is obvious that the coronal ray deflection does not need a necessary condition of the motion of a CME with a super-Alfvénic velocity. Deflection is observed also for the slow CMEs as observed in the event of 13 January 2002. In this event, the speed of the frontal structure of the CME ($v_{CMEf}$) precisely coincides with the speed of the time profile front ($v_f$), while the speed of the CME core ($v_{CMEc}$) precisely coincides with the speed of the time profile maximum $v_M$. This may be a nice matching of the observed speeds, however, of course we observe only the projection of the real speed onto the sky plane. Unknown projection effects make the real radial speed uncertain at least up to a factor of two.

We refer onwards the CMEs with a speed presumably higher than the local Alfvén speed as fast CMEs (Figure 12), and the CMEs with a speed lower than the local Alfvén speed as slow CMEs (Figure 11). However, the local value of the Alfvén speed cannot be known very precisely. It ranges from 100 km s$^{-1}$ to 2000 km s$^{-1}$ between 3.0 $R_\odot$ and 6.0 $R_\odot$ in the corona depending on the magnetic configuration of the region (Evans et al., 2008). The rather low speed of propagation of the magnetic field changes in the solar corona may result in the formation of visible internal structure of a CME. While the bright core corresponds to the central part of a flux rope containing the filament material, the frontal structure is the boundary of the flux rope magnetic field that moves at Alfvén speed. However, there may also be the other causes for the formation of the frontal part of a CME (Filippov, 1996). In slow CMEs, the frontal structure can be far away from the bright core, as it is observed in the event of 5 March 2000. In fast CMEs, the core moves close to the frontal structure, as seen in the event of 2 March 2002.

We tried to measure the speed of the disturbance propagation along rays in two spatial intervals in the best possible way. Speeds $v_{f1}$, $v_{M1}$ in Table 1 correspond to a shorter distance from the Sun, while speeds $v_{f2}$, $v_{M2}$ are measured within a more distant interval. The results are rather contradictory. Only in one event of 5 March 2000, we see deceleration of both the front and the peak of the curve. The event of 22 August 1996 shows acceleration, and the situation is much entangled on 2 March 2002. The front of the deflection of the Southern ray accelerates, while the peak decelerates. The southern ray has a constant speed for the deflection front and decelerating peak. Probably, the low image cadence does not allow us to make a definite conclusion about acceleration or deceleration of the kink propagation in this event. It is also worthwhile to note that the speeds of the peaks of Southern and Northern rays on 2 March 2002 are different, while the speeds of the fronts are nearly the same. The most likely reason is that the speed of the front of Southern ray is underestimated due to the influence of another faint and slow CME that appears 90 min before the main CME at nearly the same position angle, only 10° closer to the equator. The influence of the first CME may cause small deflection of the Southern ray that results in widening of the frontal parts of the deflection curves at smaller distances. A comparison of Figure 6 and Figure 7 shows that the deflection of the Northern ray starts a few tens of minutes later than the deflection of the Southern ray, while the angular distance of the former from the CME is approximately the same or even a little smaller. This means that the Northern ray is not located in the same meridian plane as the CME, and we see more distant parts of the ray due to the projection onto the sky plane. The projection effect may also reduce the speed of the kink propagation, which due to the projection onto the sky plane becomes lower than the speed of the CME core.

In four rays, we see the speed of the front $v_f$ is greater than the speed of the time profile maximum $v_M$. This corresponds to slow CMEs that travel with a speed $v_c$ lower than the speed $v_a$ of the magnetic field propagation in the coronal plasma. The speed of the front of the Southern ray on 2 March 2002 is possibly underestimated due to influence of the other CME. In the event of 5 March 2000, the values of $v_f$ and $v_M$ are approximately equal and deceleration is observed. One might expect that this event is a candidate for manifestation of the shock wave propagation. However, the value of the kink speed about 350 km s$^{-1}$ is not high enough, as well as the speed of the bright core of the CME is also only 200 km s$^{-1}$. Tripathi and Raouafi (2007) measured the speed of the propagating kink in this event by tracking the boundary between the bright and dark features from the running difference images and found it to be about 260 km s$^{-1}$. Nevertheless, they concluded that the streamer deflection is highly likely to be due to a CME-driven shock wave.

We conclude that all studied events do not correspond to expected pattern of the shock wave propagation in the corona. The coronal ray deflection that is clearly visible in all of them can be interpreted as the influence of the magnetic field of a moving flux rope related to a CME. When CMEs travel with the speed comparable with Alfvén speed, the sphere of their magnetic influence is limited by the speed of propagation of magnetic field changes. This volume spreads and moves along with the CME core, as shown in Figure 10. Coronal rays within the sphere of the flux rope influence are deviated. Their evolution also looks like the propagation of kink perturbations along these rays.

## 5. Conclusions

We have analyzed five events of prominent coronal ray deflections by the CMEs. We have measured the speed of the kink disturbance propagation along the rays in two spatial intervals and compared it with the speed of the frontal structure and the core of CMEs. Streamer deflections were observed for cases with both fast and slow CMEs. We did not find strong evidence of shock wave propagation in the corona. The observed coronal ray deflections can be interpreted as the influence of the magnetic field of a moving flux rope related to a CME. A simple model of radial coronal magnetic field was considered. The motion of a large-scale flux rope away from the Sun creates changes in the structure of surrounding field lines, which are similar to the kink propagation along coronal rays. The retardation of the potential is taken into account since the flux rope moves at high speed comparable with the Alfvén speed. It is expected that fast CMEs drive shocks in the corona. Distant streamer deflections are often assumed to be indirect evidence of shocks. However, there are other reasons for the streamer deflections, and at least many of them are not deflected by the shocks. Careful analysis of these events could give us valuable information about coronal plasma, for example, an estimation of Alfvén speed.

## Acknowledgements

The authors would like to thank the anonymous referee for constructive comments. This work was supported in part by the Russian Foundation for Basic Research (grants 09-02-00080 and 09-02-92626) and in part by the Department of Science and Technology, Ministry of Science and Technology of India (INT/RFBR/P-38).